# BAlN alloy for enhanced two-dimensional electron gas characteristics of GaN-based high electron mobility transistor

Rongyu Lin[1], Xinwei Liu[1], Kaikai Liu[1], Yi Lu[1], Xinke Liu[2*], and Xiaohang Li[1*]

[1]*King Abdullah University of Science and Technology (KAUST), Advanced Semiconductor Laboratory, Thuwal 23955-6900, Saudi Arabia*

[2]*College of Materials Science and Engineering, College of Electronics and Information Engineering, Guangdong Research Center for Interfacial Engineering of Functional Materials, Shenzhen Key Laboratory of Special Functional Materials, Chinese Engineering and Research Institute of Microelectronics, Shenzhen University, Shenzhen 518060, China*

\*Authors to whom correspondence should be addressed:

*xkliu@szu.edu.cn; xiaohang.li@kaust.edu.sa

**Abstract:**

The emerging wide bandgap BAlN alloys have potentials for improved III-nitride power devices including high electron mobility transistor (HEMT). Yet few relevant studies have been carried. In this work, we have investigated the use of the $B_{0.14}Al_{0.86}N$ alloy as part or entirety of the interlayer between the GaN buffer and the AlGaN barrier in the conventional GaN-based high electron mobility transistor (HEMT). The numerical results show considerable improvement of the two-dimensional electron gas (2DEG) concentration with small 2DEG leakage into the ternary layer by replacing the conventional AlN interlayer by either the $B_{0.14}Al_{0.86}N$ interlayer or the $B_{0.14}Al_{0.86}N$/AlN hybrid interlayer. Consequently, the transfer characteristics can be improved. The saturation current can be enhanced as well. For instance, the saturation currents for HEMTs with the 0.5 nm $B_{0.14}Al_{0.86}N$/0.5 nm AlN hybrid interlayer and the 1 nm $B_{0.14}Al_{0.86}N$ interlayer are 5.8% and 2.2% higher than that for the AlN interlayer when $V_{GS}-V_{th}= +3$ V.

# 1. Introduction

In the past decades, III-nitride semiconductors especially AlGaN, InGaN, and GaN have been widely employed for optical and electronic devices due to superior properties such as suitable bandgap, high saturation velocity, high breakdown field, and high chemical and thermal stability. Invented in 1993 based on the two-dimensional electron gas (2DEG) at the AlGaN/GaN heterointerface induced by the net polarization charge, the GaN-based high electron mobility transistor (HEMT) has attracted enormous research interest due to its vast high power and high speed applications [1]. Since then, substantial efforts have been made to improve the GaN-based HEMT performance [2].

In general, a higher Al content of the AlGaN barrier layer can induce larger net polarization charge at the AlGaN/GaN heterointerface, resulting in higher 2DEG concentration in the conventional AlGaN/GaN HEMT structure. However, the electron mobility can be compromised due to the scattering caused by the alloy disorder [3]. Additionally, further increasing the Al content of the AlGaN barrier layer could degrade the interface quality due to the larger lattice mismatch between the AlGaN barrier layer and the GaN buffer layer [3].

To overcome the issues, the insertion of a thin ($\leq 2$ nm) AlN interlayer between the AlGaN barrier layer and the GaN channel layer has been proposed and implemented, partly because the binary AlN does not cause alloy scattering [4, 5]. A thicker AlN interlayer is not feasible, as the AlN interlayer could relax due to the large lattice mismatch with the GaN buffer layer. Furthermore, the AlN interlayer can elevate the barrier height for the 2DEG due to its higher conduction band edge than that of the AlGaN barrier. Since the penetration of the electron wave function into a given barrier is in inverse proportion to the barrier height, the AlN interlayer could suppress the 2DEG leakage into the AlGaN barrier, further suppressing the alloy scattering [6]. Moreover, the net polarization charge at the AlN/GaN heterointerface is greater than that of the AlGaN/GaN heterointerface, which induces higher 2DEG concentration. Combining the three benefits, the AlN interlayer has been employed to minimize the alloy scattering and enhance the 2DEG concentration [7].

The wurtzite BAlN alloys with large bandgap have application potentials for III-nitride devices. Recently, the meaningful progress has been made in studying epitaxy and properties of BAlN thin films and heterojunctions. The growth of highly reflective $B_xAl_{1-x}N$/AlN distributed Bragg reflector (DBR) was conducted by metalorganic vapor phase epitaxy (MOVPE) [8]. The growth of five-period $B_xAl_{1-x}N$/AlN heterostructures with the boron content of 11% has been demonstrated [9]. Recently, researchers have performed the growth of single-crystalline BAlN layers with relatively high boron composition by MOVPE [10-12]. Moreover, the lattice constant and polarization properties of the BAlN alloys have been studied, in which large variations of those important quantities were found [13-15]. Furthermore, the band alignments of the BAlN/(Al)GaN heterojunctions have been determined, showing an extremely large conduction band offset between BAlN and (Al)GaN due to the type-II nature [16, 17]. Despite those efforts, however, there have been few device research reports about the incorporation of the BAlN alloys in the technically-important HEMTs.

In this work, we propose to incorporate the BAlN alloy into the GaN HEMTs by leveraging its unique band alignment and polarization properties. Specifically, a thin BAlN layer is employed as the whole interlayer or as part of the interlayer in the conventional AlGaN/GaN-based HEMTs to provide considerably larger polarization charge at the heterointerface and thus 2DEG concentrations; while to effectively suppress the electron leakage into the AlGaN barrier layer through larger conduction band offset than the conventional AlN interlayer.

## 2. Impact of interlayers on 2DEG characteristics

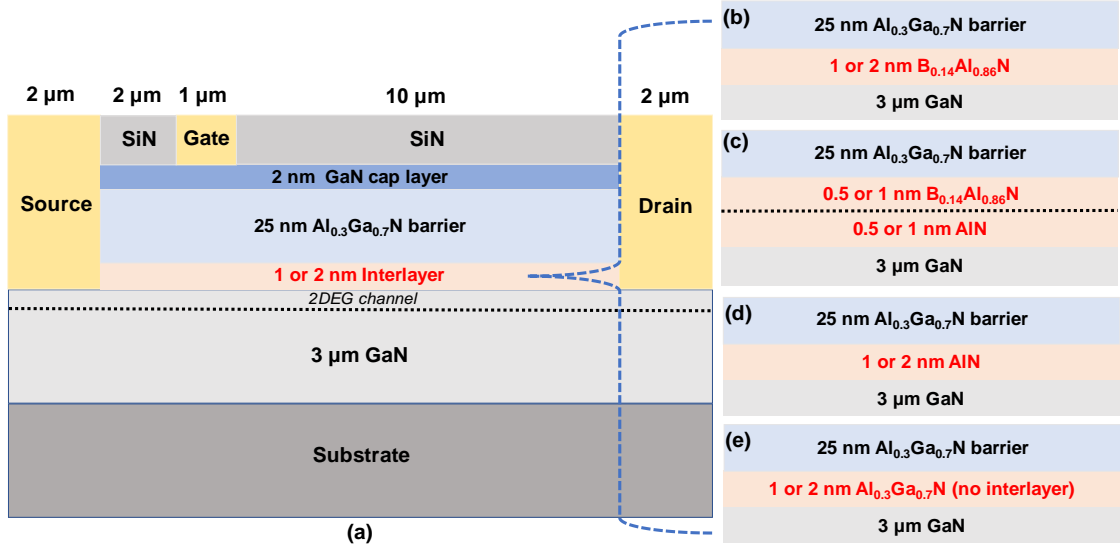

Fig. 1.(a) Cross-sectional schematics of the HEMT structures with different interlayer configurations: (b) 1 or 2 nm $B_{0.14}Al_{0.86}N$, (c) 0.5 nm $B_{0.14}Al_{0.86}N$/0.5 nm AlN or 1 nm $B_{0.14}Al_{0.86}N$/1 nm AlN, (d) 1 or 2 nm AlN, (e) 1 or 2 nm $Al_{0.3}Ga_{0.7}N$ (i.e. no interlayer).

The simulation structures are constructed based on the conventional wurtzite III-polar GaN-based HEMT comprising, a fully-relaxed GaN buffer layer, a 1 or 2 nm interlayer, and an $Al_{0.3}Ga_{0.7}N$ barrier layer shown in Fig. 1(a) [18-20]. For the interlayers, various designs under consideration include: 1 nm $B_{0.14}Al_{0.86}N$ interlayer, 2 nm $B_{0.14}Al_{0.86}N$ interlayer, 0.5 nm $B_{0.14}Al_{0.86}N$/0.5 nm AlN hybrid interlayer, 1 nm $B_{0.14}Al_{0.86}N$/1 nm AlN hybrid interlayer, 1 nm AlN interlayer, and 2 nm AlN interlayer, as shown in Fig. 1(b)-(e). It should be noted that for the structures without the interlayer, we increase the $Al_{0.3}Ga_{0.7}N$ barrier thickness from 25 nm to 26 and 27 nm corresponding to the 1 and 2 nm $Al_{0.3}Ga_{0.7}N$ interlayers, respectively, in order to keep the total thickness fixed for fair comparison. The B-content of 14% is chosen, i.e. the $B_{0.14}Al_{0.86}N$ alloy, because its band alignment with (Al)GaN could be promising for the enhanced GaN-based HEMT and it has been realized epitaxially [16, 17]. The maximum thickness of the interlayer is limited at 2 nm as larger thicknesses could lead to lattice relaxation given the large lattice mismatch between the interlayer and the buffer layer. The entire HEMT structure is assumed to be fully strained to the GaN buffer layer. The $B_{0.14}Al_{0.86}N$/AlN heterojunction where AlN situates between $B_{0.14}Al_{0.86}N$ and GaN is employed to minimize potential alloy scattering. The dielectric $SiN_x$ is applied for surface passivation. The length of the Schottky gate, the gate-to-source spacing, and the gate-to-drain spacing are set according to the typical values of the GaN-based HEMT shown in Fig. 1(a).

The simulation employs APSYS developed by Crosslight Inc [21]. The electrical properties of the structures are performed by solving the Poisson's equation and the continuity equation. The transport model of electrons and holes including drift and diffusion are considered. The formation of the 2DEG amid the HEMT structures are mainly related to the heterointerface polarization difference and the band offset. The lattice constants, polarization parameters, and band offsets of the involved materials are from Refs [13, 16, 17, 22], from which it can be deduced that the $B_{0.14}Al_{0.86}N$/GaN heterojunction exhibits significantly larger conduction band offset and heterointerface polarization difference as compared to the conventional AlN/GaN and $Al_{0.3}Ga_{0.7}N$/GaN heterojunctions, potentially leading to much higher 2DEG concentration and less 2DEG leakage from the channel into the barrier.

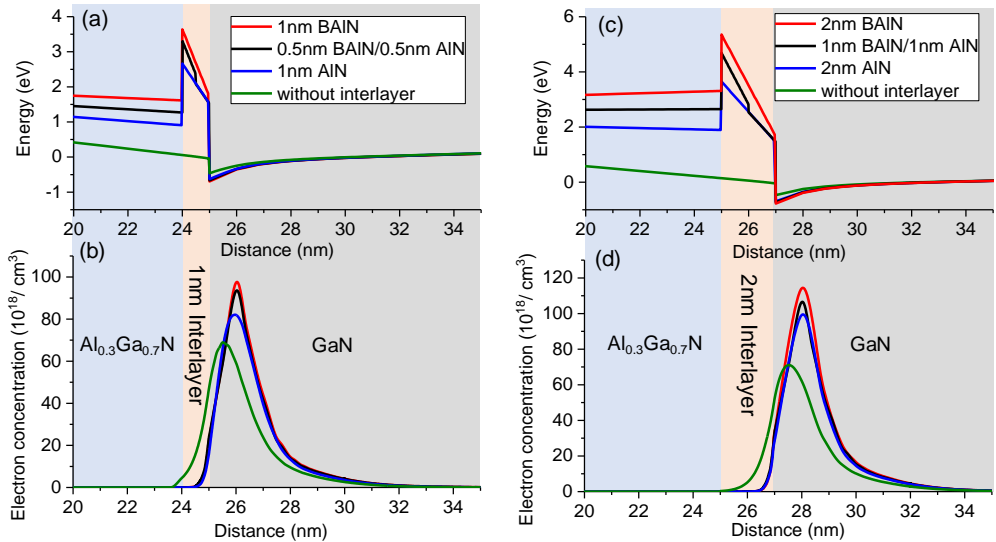

Fig. 2. (a) Band diagram and (b) electron concentration of the HEMT structures with and without the 1 nm interlayers. (c) Band diagram and (d) electron concentration of the HEMT structures with and without the 2 nm interlayers.

The conduction band diagrams and the polarization-induced electron concentration of the HEMT structures with and without the 1 and 2 nm interlayers under zero bias are plotted in Fig. 2(a)-(d). Due to larger polarization sheet charge induced by $B_{0.14}Al_{0.86}N$, the conduction bands of the GaN buffer layer bend down the most below the Fermi level at the interlayer/buffer interface for the 1 and 2 nm $B_{0.14}Al_{0.86}N$ interlayers om Fig. 2(a) and (c), respectively, followed by the hybrid interlayers and the AlN interlayers. The structures without the interlayers have the smallest conduction band bending. As the increased band bending below the Fermi level induces more electrons at the channel, the use of the $B_{0.14}Al_{0.86}N$ interlayers leads to the highest concentration of electrons, followed by the hybrid interlayers, the AlN interlayers, and the structures without the interlayers shown in Fig. 2(b) and (d).

The corresponding 2DEG sheet densities of different structures are shown by red squares in Fig. 3. The structures with the 1 nm $B_{0.14}Al_{0.86}N$ interlayer and the 1 nm hybrid interlayer have 2DEG sheet densities of $1.834 \times 10^{13}$ and $1.710 \times 10^{13}$ /cm$^2$, which are 11.6% and 4.0% higher than $1.644 \times 10^{13}$ /cm$^2$ for the 1 nm AlN interlayer, respectively. Additionally, the 2DEG sheet densities of the structures with the 2 nm $B_{0.14}Al_{0.86}N$ and hybrid interlayers are $2.120 \times 10^{13}$ and $2.001 \times 10^{13}$ /cm$^2$, which are 13.6% and 7.2% larger than $1.866 \times 10^{13}$ /cm$^2$ for the 2 nm AlN interlayer. The structures without the interlayers, i.e. 1 nm $Al_{0.3}Ga_{0.7}N$ and 2 nm $Al_{0.3}Ga_{0.7}N$ in Fig. 3, have the lowest 2DEG sheet densities of $1.532 \times 10^{13}$ and $1.517 \times 10^{13}$ /cm$^2$, respectively.

Though the $B_{0.14}Al_{0.86}N$ interlayer and the hybrid interlayer have shown considerably higher 2DEG sheet density, the leakage of the 2DEG into the $B_{0.14}Al_{0.86}N$ ternary alloy could lead to scattering induced by the alloy disorder, that could compromise the 2DEG mobility and the device characteristics [23]. Meanwhile, Fig. 2(a) and (c) show the structures with the 1 and 2 nm $B_{0.14}Al_{0.86}N$ interlayers exhibit the highest conduction band edge in the respective figures contributing to large effective conduction band offsets of 4.33 and 6.13 eV between the interlayer and the channel, respectively, which is desirable for the suppressed 2DEG leakage. The $B_{0.14}Al_{0.86}N$/AlN hybrid interlayers result in slightly lower effective conduction band offset but still higher than that by the conventional AlN interlayers in Fig. 2(a) and (c). The structures without the interlayer have drastically lower conduction band edge and hence the smallest effective conduction band offset.

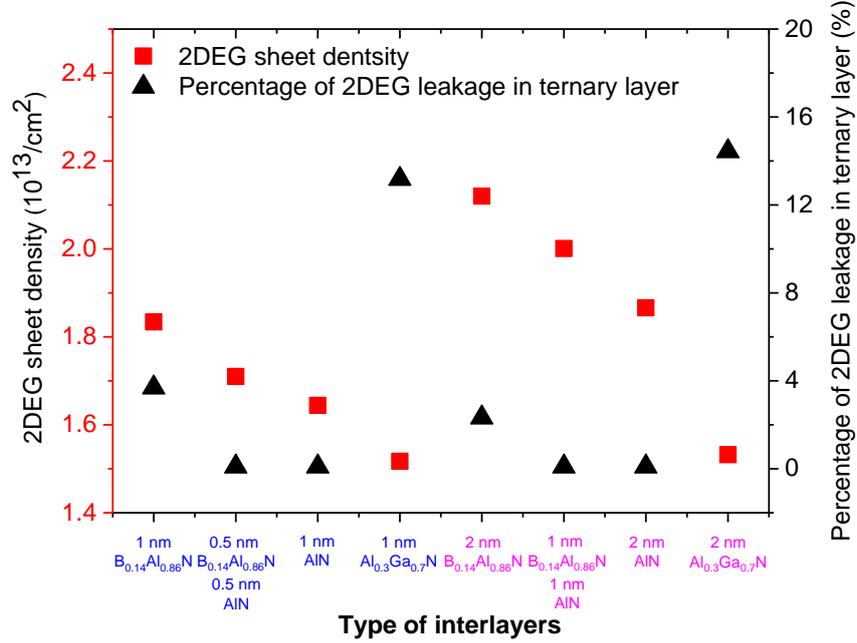

Fig. 3. The 2DEG sheet density (red) and the percentage of 2DEG leakage in the ternary layer (black) of the HEMT structures with different configurations of 1 nm interlayers (blue axis) and 2 nm interlayers (purple axis).

Fig. 2(b) and (d) manifest the spatial distribution of the 2DEG in the vicinity of the GaN channel. Majority of the 2DEG of all the structures situates amid the GaN channel, though the structures without the interlayers have the largest leakage of the 2DEG in the $Al_{0.3}Ga_{0.7}N$ ternary layer. To quantify the leakage into the ternary layer, i.e. $Al_{0.3}Ga_{0.7}N$ or $B_{0.14}Al_{0.86}N$, the percentage of the 2DEG leakage to the ternary layer is defined as follows and the corresponding values are shown by the black triangles in Fig. 3.

$$Percentage\ of\ 2DEG\ leakage = \frac{2DEG\ in\ ternary\ layer}{Total\ 2DEG} \qquad (1)$$

Large percentages of 13.2% and 14.4% are found for the 1 and 2 nm $Al_{0.3}Ga_{0.7}N$ (i.e. no interlayer), which can lead to severe alloy scattering. The percentages for the 1 and 2 nm AlN interlayers are less than 0.1% which show effective leakage suppression. The small percentages less than 0.1% are also found for the 1 and 2 nm $B_{0.14}Al_{0.86}N$/AlN hybrid interlayers, indicating excellent suppression with larger 2DEG sheet densities as compared with the AlN interlayers. For the 1 and 2 nm $B_{0.14}Al_{0.86}N$ interlayers with the largest 2DEG sheet densities, the relatively small percentages of 3.7% and 2.3% are found.

## 3. HEMT device characteristics

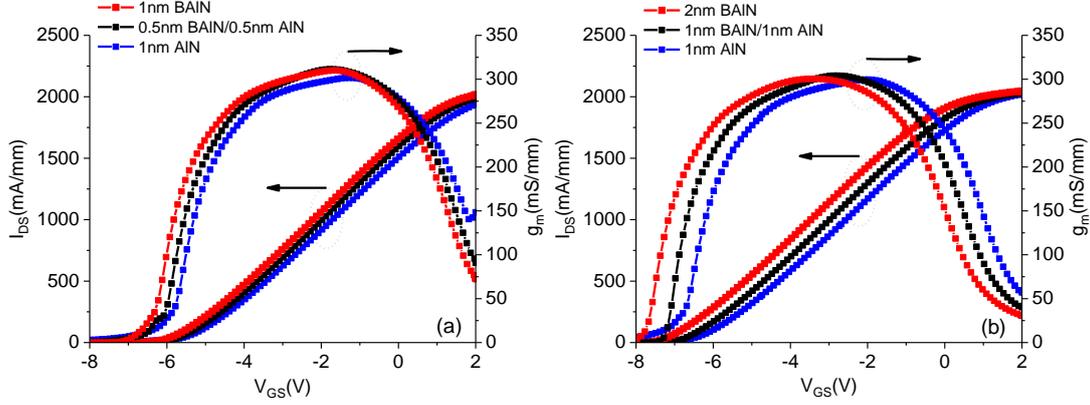

Fig. 4. Transfer characteristics under $V_{DS}$ = 10 V of the HEMT structures comprising interlayers of (a) 1 nm $B_{0.14}Al_{0.86}N$, 0.5 nm $B_{0.14}Al_{0.86}N$/0.5 nm AlN, and 1 nm AlN; and (b) 2 nm $B_{0.14}Al_{0.86}N$, 1 nm $B_{0.14}Al_{0.86}N$ /1 nm AlN, and 2 nm AlN.

The HEMT device characteristics are investigated subsequent to the study of the 2DEG characteristics. From Fig. 4, the threshold voltages under $V_{DS}$ = 10 V for the HEMTs with the interlayers of 1 nm $B_{0.14}Al_{0.86}N$, 0.5 nm $B_{0.14}Al_{0.86}N$/0.5 nm AlN, and 1 nm AlN are -5.43, -5.18, and -5.03 V, respectively, which are inversely proportional to the 2DEG sheet density in Fig. 3. Similarly, for the 2 nm interlayer designs, the threshold voltages are -6.79, -6.27, and -5.89 V for the 2 nm $B_{0.14}Al_{0.86}N$, 1 nm $B_{0.14}Al_{0.86}N$/1 nm AlN, and 2 nm AlN interlayers. The overall decreased threshold voltages for the 2 nm interlayer designs are attributed to the higher 2DEG sheet densities, in comparison with the 1 nm interlayer designs shown in Fig. 3.

Besides, the 0.5 nm $B_{0.14}Al_{0.86}N$/0.5 nm AlN hybrid interlayer exhibits the maximum transconductance $g_m$ of 311.36 mS/mm, compared with 310.00 mS/mm for the 1 nm $B_{0.14}Al_{0.86}N$ interlayer and 301.6 mS/mm for the 1 nm AlN interlayer. Since the gate length remains the same for all structures of interest, the highest maximum transconductance for the 0.5 nm $B_{0.14}Al_{0.86}N$/0.5 nm AlN hybrid interlayer indicates improved carrier transport attributed to the minimized 2DEG leakage into the ternary layer and the higher 2DEG sheet density shown in Fig. 2 and Fig. 3. Despite with the 2DEG leakage of 3.7%, the 1 nm $B_{0.14}Al_{0.86}N$ interlayer design yields larger maximum transconductance as opposed to the conventional 1 nm AlN interlayer. Also the transconductance of 310.00 mS/mm for the 1 nm $B_{0.14}Al_{0.86}N$ interlayer is barely lower than that for the 1 nm hybrid interlayer which could be attributed to the larger 2DEG sheet density. For the structures with the 2 nm interlayers, the hybrid interlayer comprising 1 nm $B_{0.14}Al_{0.86}N$/1 nm AlN shows the largest maximum transconductance of 304.13 mS/mm among the three, compared to the 2 nm $B_{0.14}Al_{0.86}N$ interlayer (300.50 mS/mm) and the 2 nm AlN interlayer (299.63 mS/mm). Again, the $B_{0.14}Al_{0.86}N$-containing interlayers lead to higher transconductance than the AlN interlayer for the 2 nm interlayer design. The larger transconductances show feasibility of employing $B_{0.14}Al_{0.86}N$ as the entirety or part of the interlayer to enhance the channel carrier transport.

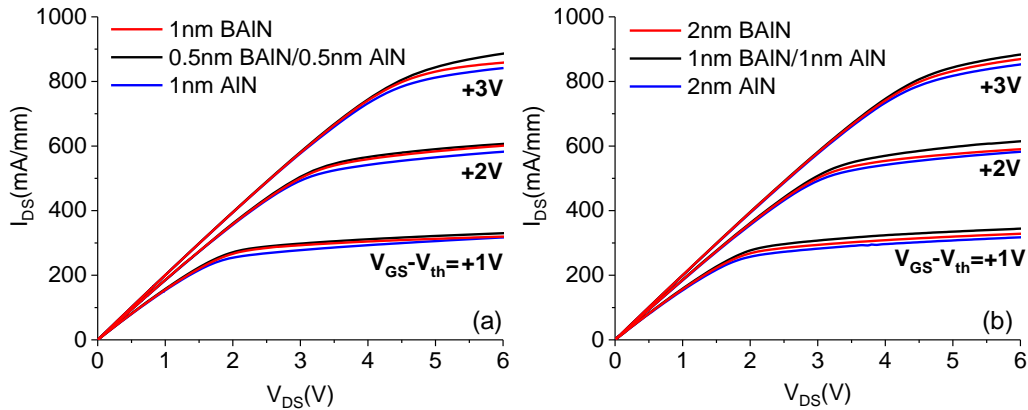

Fig. 5. The $I_{DS}$ - $V_{DS}$ characteristics of the HEMT structures comprising interlayers of (a) 1 nm $B_{0.14}Al_{0.86}N$, 0.5 nm $B_{0.14}Al_{0.86}N$/0.5 nm AlN, and 1 nm AlN; and (b) 2 nm $B_{0.14}Al_{0.86}N$, 1 nm $B_{0.14}Al_{0.86}N$ /1 nm AlN, and 2 nm AlN.

Fig. 5(a) and (b) illustrates the $I_{DS}$ - $V_{DS}$ characteristics of the HEMT structures with different 1 and 2 nm interlayers with varying $V_{GS}$-$V_{th}$ of +1, +2, and +3 V. In all cases of Fig. 5(a) and (b), the hybrid $B_{0.14}Al_{0.86}N$/AlN interlayer results in higher saturation current under the same $V_{GS}$-$V_{th}$, with the HEMTs with the $B_{0.14}Al_{0.86}N$ interlayer and the AlN interlayer being the second and the third, respectively. For instance, the saturation currents for the $B_{0.14}Al_{0.86}N$/AlN hybrid interlayer and the $B_{0.14}Al_{0.86}N$ interlayer are 5.8% and 2.2% higher than for the AlN interlayer when $V_{GS}$-$V_{th}$= +3 V in Fig. 5(a) for the 1 nm interlayers. For the 2 nm interlayers in Fig. 5(b), 3.7% and 1.9% saturation current enhancements are observed in the hybrid $B_{0.14}Al_{0.86}N$/AlN interlayer and the $B_{0.14}Al_{0.86}N$ interlayer compared with that for the AlN interlayer under same $V_{GS}$-$V_{th}$. The enhancements are consistent with the enhanced maximum transconductance in Fig. 4(a) and (b) for the HEMTs comprising the $B_{0.14}Al_{0.86}N$-based interlayers.

## 3. Conclusion

We have proposed the application of the $B_{0.14}Al_{0.86}N$ alloy for the interlayers of GaN-based HEMT device structures. The proposed structures comprise either $B_{0.14}Al_{0.86}N$ interlayers or $B_{0.14}Al_{0.86}N$/AlN hybrid interlayers of 1 or 2 nm thick. The proposed structures lead to considerable enhancement of the 2DEG sheet density as opposed to the conventional AlN interlayers and result in good 2DEG confinement with low 2DEG leakage into the ternary layer. For instance, the structures with the 1 nm $B_{0.14}Al_{0.86}N$ interlayer and the 1 nm hybrid interlayer have 2DEG sheet densities of $1.834 \times 10^{13}$ and $1.710 \times 10^{13}$ /cm$^2$, which are 11.6% and 4.0% higher than $1.644 \times 10^{13}$ /cm$^2$ for the 1 nm AlN interlayer, respectively. Additionally, the small percentages of the 2DEG leakage of less than 0.1% are found for the 1 and 2 nm $B_{0.14}Al_{0.86}N$/AlN hybrid interlayers, indicating excellent suppression with larger 2DEG sheet densities. The enhanced 2DEG characteristics thanks to the use of the $B_{0.14}Al_{0.86}N$ alloy lead to higher maximum transconductance and saturation currents than the HEMTs with the conventional AlN interlayers.

**Acknowledgments**

The KAUST authors would like to acknowledge the support of like to acknowledge the support of KAUST Baseline Fund BAS/1/1664-01-01, GCC Research Council Grant REP/1/3189-01-01, and Competitive Research Grants URF/1/3437-01-01 and URF/1/3771-01-01.